\documentclass[twocolumn,showpacs,superscriptaddress,prb,aps,floatfix]{revtex4-1}
\usepackage{graphicx}
\usepackage{rotating}
\usepackage{amsmath}
\usepackage[usenames,dvipsnames]{color}
\usepackage{float}
\usepackage{lineno}

\makeatletter

\newcommand{\Rmnum}[1]{\expandafter\@slowromancap\romannumeral #1@}
\makeatletter

\hfuzz=\maxdimen
\begin{document}

\title{Structural determination and electronic properties of 4$d$ perovskite SrPdO$_3$}

\author{Jiangang He}
\affiliation{University of Vienna, Faculty of Physics and Center for Computational
Materials Science, Vienna, Austria}
\affiliation{School of Applied \& Engineering Physics, Cornell University,
Ithaca, NY 14853, USA}

\author{Cesare Franchini}
\affiliation{University of Vienna, Faculty of Physics and Center for Computational
Materials Science, Vienna, Austria}

\date{\today}
\pacs{61.50.Ah, 71.15.Mb, 71.20.-b, 63.20.-e, 75.10.-b,}

\begin{abstract}
The structure and ground state electronic structure of the recently synthesized SrPdO$_3$ perovskite
[A. Galal {\em et al.}, J. Power Sources, {\bf 195}, 3806 (2010)] have been studied by means of screened
hybrid functional and the GW approximation with the inclusion of electron-hole interaction within the test-charge/test-charge 
scheme. By conducting a structural search based on lattice dynamics and group
theoretical method we identify the orthorhombic phase with $P_{nma}$ space group as the most stable crystal
structure. The phase transition from the ideal cubic perovskite structure to the $P_{nma}$ one is explained
in terms of the simultaneous stabilization of the antiferrodistortive phonon modes $R_4^+$ and $M_3^+$.
Our results indicate that SrPdO$_3$ exhibits an insulating ground state, substantiated by a GW$_0$ gap of about 1.1 eV.
Spin polarized calculations suggests that SrPdO$_3$ adopts a low spin
state ($t_{2g}^{\uparrow\downarrow\uparrow\downarrow\uparrow\downarrow}e_g^0$), and is expected to exhibit
spin excitations and spin state crossovers at finite temperature, analogous to the case of 3$d$ isoelectronic
LaCoO$_3$. This would provide a new playground for the study of spin state transitions in 4$d$ oxides and
new opportunity to design multifunctional materials based on 4$d$ $P_{nma}$ building block.
\end{abstract}

\maketitle


\section{INTRODUCTION}
Owing to the wide range of physical and chemical properties, transition metal (TM)  oxides perovskites
(chemical formula $AB$O$_3$) exhibit
a large variety of functional behaviors (Ref.\onlinecite{Vrejoiu08}) and find applications in many technologically relevant fields 
including superconductivity,\cite{maeno,Habermeier07} multiferroicity,\cite{kimura,Ramesh07,Stroppa12} solid oxide fuel
cells,\cite{adler, Suntivich11} catalysis,\cite{Misono05} and thermoelectricity.\cite{Ohta07}
In the last decade, 4$d$ TM perovskites have attracted increasing attention motivated by
the numerous intriguing (and to some extent unexpected) properties originating from the delicate balance between the
larger spatial extension of the 4$d$ manifold, the reduced Coulomb repulsion and exchange energy, and the increased
tendency towards O-$2p$/TM-$4d$ hybridization.\cite{Lee03} As compared to the most widely studied 3$d$ TM oxides, 
4$d$ TM perovskites offer a novel physical arena to study the interactions
among spin, orbital, electron, and lattice degrees of freedom.
Prime examples of these anomalous behaviors (non-Fermi liquid,
ferromagnetism, half-metallicity, high-temperature anti-ferromagnetism, high conductivity, etc.) were recognized
in ruthanates,\cite{allen,kostic,Rondinelli08} molybdates\cite{nagai} and technetiates.\cite{efrain,franchini11,Middey12,Mravlje12}
In addition to these fundamental issues, the growing concerns about global warming have boosted an intense
research on energy production processes based on perovskites. In particular, noble metal incorporated perovskites
represent a valuable alternative as catalysts since they have a high thermal stability and
their elevated compositional flexibility along with the presence of chemically active structural defects and localized
electronic states offer tremendous tailoring possibilities for designing more efficient catalytic
processes.\cite{Tanaka02,Sun10,Suntivich11}
Palladium oxides is one of the most useful catalyst in oxidation reaction\cite{yeung99,meng11} and the incorporation of Pd
in perovskite structure turned out to efficiently suppress the inconvenient clustering of Pd particles, thereby improving the
catalytic activity.\cite{Tanaka02}

Very recently, SrPdO$_3$ perovskite with a high catalytic activity toward hydrogen evolution reaction was successfully
synthesized for the first time by Galal {\em et. al.} by a citrate method.\cite{galal}
These authors have also shown, that the Curie-Weiss plot derived from the evolution of the molar magnetic
susceptibility as a function of temperatures (for temperatures higher than 280 K),  exhibits an 
antiferromagnetic (AFM) character, with a magnetic
ordering temperature of about 370 K and a surprisingly large calculated magnetic moment of 5.7 $\mu_{\rm B}$, suggestive
of a high-spin (HS) configuration ($t_{2g}^{\uparrow\downarrow\uparrow\uparrow}e_g^{\uparrow\uparrow}$) of the Pd$^{4+}$
ion.\cite{galal} Considering that the spin-state of $d^6$ perovskites such as LaCoO$_3$ represents a great challenge for
both theory and experiment and have been the source of much controversy\cite{Goodenough01,Rao04} these findings
should be taken with a certain caution. At low temperature, in fact, LaCoO$_3$ (isoelectronic to SrPdO$_3$) posses
a low-spin (LS) configuration ($t_{2g}^{\uparrow\downarrow\uparrow\downarrow\uparrow\downarrow}e_g^0$) and a diamagnetic
semiconducting character.\cite{saitoh,tokura} As the temperature is elevated LaCoO$_3$ exhibits an intricate magnetic
phase diagram characterized by an intermediated LS-HS state, possibly associated with a LS-HS crossover,\cite{Raccah67}
whose detailed nature remains highly debated.\cite{Hsu10}
The HS state is usually favored when the crystal field splitting energy is smaller than the Coulomb repulsive energy 
among electrons. As 4$d$ ions generally have larger crystal field splitting energy and smaller Coulomb repulsive
energy than 3$d$ ions, it is expected that 4$d$ TM oxides such as SrPdO$_3$ would prefer a LS configuration, subjected to
spin-state crossover driven by thermal excitations.

The electronic structure and the detailed structural properties of SrPdO$_3$ remain up to date to be determined.
The scope of the present work is to address these issues by means of an accurate {\em ab initio} analysis
performed in the framework of hybrid density function theory and GW approximation. In particular, we have
determined the most favorable crystal symmetry and structural distortions on the basis of a careful analysis of
the phonon instabilities starting from the ideal cubic perovskite structure.\cite{parlinski00,lebedev}
After describing the computational set-up (Sec. \ref{sec:cm}), we will discuss the results of our structural
search (Secs. \ref{sec:ph} and \ref{sec:st}). The electronic properties of the most stable structure will be presented
in Sec. \ref{sec:el}. The most important outcomes of our study are summarized in the last section.

\section{COMPUTATIONAL METHOD}
\label{sec:cm}

All calculations were performed with the Vienna \emph{ab intio} Simulation Package (VASP).\cite{gk1,gk2} 
The electron and ion interaction was described within the projector augmented wave (PAW) method.\cite{blochl,gk4}
The plane wave basis set cutoff energy was set to 400 eV and the {\it k}-space integration was done by adopting
a 8$\times$8$\times$8 Monkhorst-Pack grid for the cubic $P_{m\bar{3}m}$ structure and the same density was kept 
for all other considered phases.
All structures were fully relaxed  (including unit cell shape, volume, and internal atomic positions) 
by using the screened hybrid density functional scheme of Heyd-Scuseria-Ernzerhof (HSE06, Ref.\onlinecite{heyd})
until the Hellmann-Feynman forces acting on each atom were smaller than 0.01 eV/\AA.

The phonon dispersion of ideal cubic and ground state $P_{nma}$ perovskite structures 
were calculated by means of the finite difference method as implemented in PHONOPY,\cite{phonon1,phonopy}
by adopting a 2$\times2\times2$ supercell for both the $P_{m\bar{3}m}$ and $P_{nma}$ phase. 
The cubic phonon dispersion was computed at HSE06 level, whereas for the larger $P_{nma}$ supercell due to the exceedingly high 
computational cost we used the Perdew-Becke-Ernzerhof (PBE, Ref. \onlinecite{PBE}) functional and the results were cross-checked by
LDA (Ref. \onlinecite{LDA}) and  PBEsol (Ref. \onlinecite{PBEsol}) functionals.

The ground state electronic structure of $P_{nma}$ structure was studied by using HSE with different values of the Hartree-Fock exchange mixing parameter $\alpha$
and the GW approximation (Ref. \onlinecite{hedin}) within the plasmon-pole model as implemented in the VASP code.\cite{Shishkin1,Shishkin2}
The GW calculations were performed partially self-consistently, by keeping the screened Coulomb interaction W fixed to the initial W$_0$ value and iterating only the Green function G
until the quasi-particle energies were converged (this was achieved after four iterations). This partial self-consistent procedure
is usually refereed to as GW$_0$. Since the results of GW$_0$ calculations critically depends on the initial 
orbitals used to construct the screening properties and on the treatment of excitonic effects,\cite{Fuchs07, Paier08, Franchini10}
we have tested both PBE and HSE starting orbitals and for selected cases we have incorporated
electrostatic interactions between electrons and hole within the test-charge/test-charge (TC-TC) scheme.\cite{Bruneval05}
In the GW calculations we have included 320 bands, and the TCTC screening was computed for 80 bands around the 
Fermi energy.
 
As already discussed in the introduction the determination of the spin state of d$^6$ perovskites is affected
by large confusion due to the approximation involved in {\em ab initio} schemes and by the difficulty in
reliably extracting the spin state from experimental data. We have scrutinized different magnetic ordering
(AFM-A, AFM-C, AFM-G and ferromagnetic, see Ref. \onlinecite{he12}) and, in line with the results obtained by HSE for LaCoO$_3$,\cite{he12}
we found that all structures are predicted to adopt a LS configuration.

\section{RESULTS AND DISCUSSION}
\label{sec:res}

\begin{figure}
\includegraphics[clip,width=0.49\textwidth]{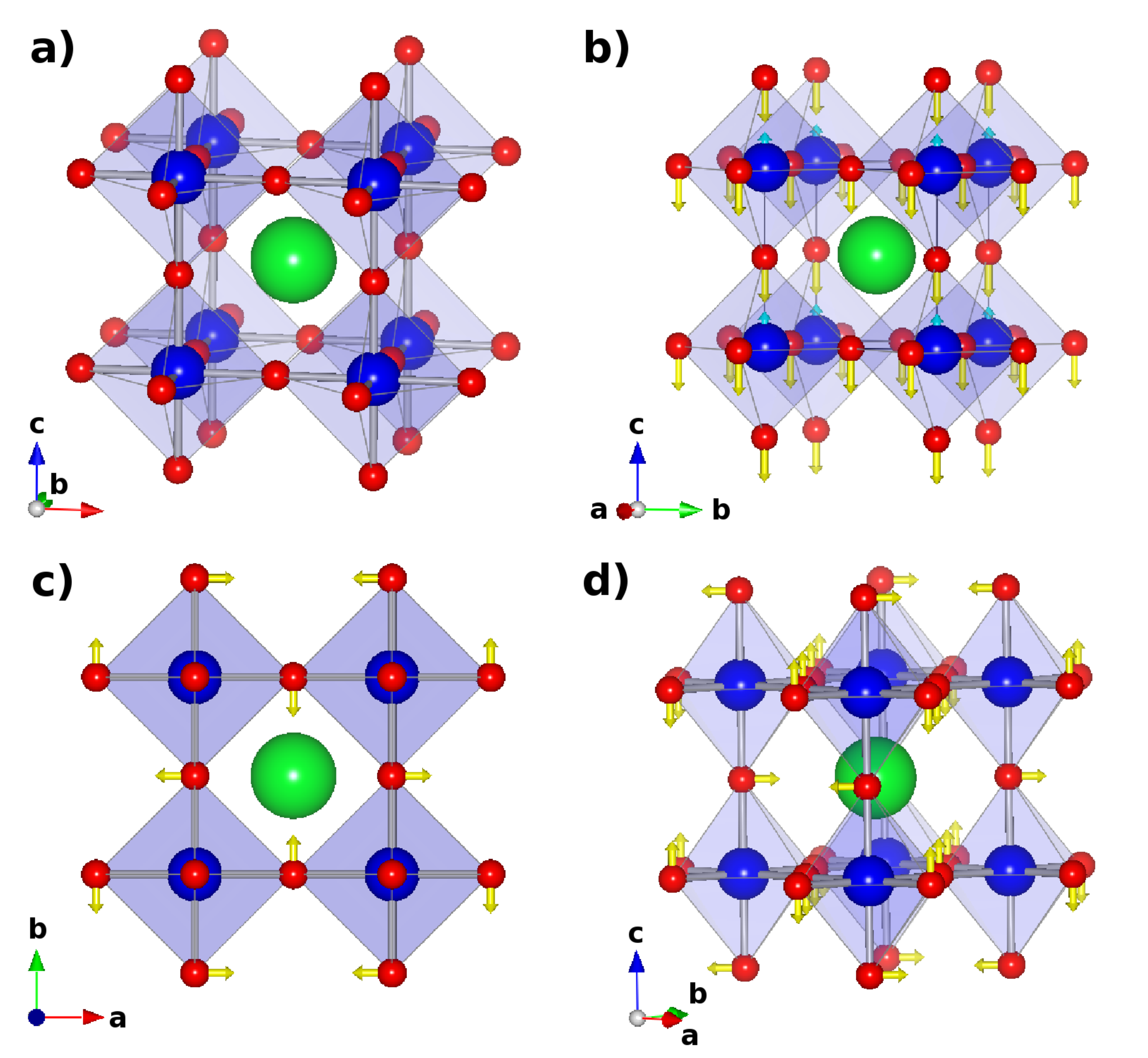}
\caption
{(Color online) Structural model and relevant phonon displacements.
(a) Structure of ideal cubic ($P_{m\bar{3}m}$) SrPdO$_3$.
In (b), (c), and (d) it is shown an illustration of the atomic
displacements associated with the symmetry modes $\Gamma_4^-$, $M_3^+$, and $R_4^+$, respectively.
Large (green), medium-size (blue) and small (red) spheres indicate Sr, Pd, and O atoms, respectively
The arrows represent the displacement of oxygen and Pd atom.
}
\label{fig:1}
\end{figure}

In the first part of this section we determine the unstable phonon modes of the high symmetry ($P_{m\bar{3}m}$, No. 221)
cubic perovskite structure, which will serve as guidelines to setup the strategy for the structural search. This section ends
with the discussion of the ground state electronic properties of the most stable phase of SrPbO$_3$.

\subsection{Phonon dispersion of cubic phase}
\label{sec:ph}

The ideal $AB$O$_3$ perovskite has cubic symmetry (space group $P_{m\bar{3}m}$, No. 221), and can be structurally interpreted as a
framework of corner-sharing $B$O$_6$ octahedra with the $A$ cations located at the 12-fold-coordinated 
intra-octahedra voids.
The stability and distortion of perovskite crystal structures are often discussed in terms of the
Goldschmidt’s tolerance factor $t$, $t = (r_A + r_O)/\sqrt{2}(r_B + r_O)$, where $r_A$, $r_B$, and $r_O$ are the
ionic radii of $A$, $B$, and O ions, respectively.\cite{Goldschmidt,Shannon} $t \approx 1$ represents the ideal conditions
upon which the perovskite structure assumes its ideal cubic symmetry.
Depending on whether the tolerance factor $t$ is greater or less than 1.0, different kind of distorted structural variants
are formed: hexagonal (face-sharing $B$O$_6$ octahedra) for $t > 1$,  whereas for $t < 0.9$ cooperative rotations of the
octahedra yield lower symmetry variants such as tetragonal, orthorhombic, rhombohedral etc.\cite{Goldschmidt}
As SrPdO$_3$ has $t=0.905$ it is expected that the cubic structure is dynamical unstable at low
temperature, which could be stabilized throughout suitable rotations and tilting of the octahedra network.

\begin{figure}
\includegraphics[clip,width=0.49\textwidth]{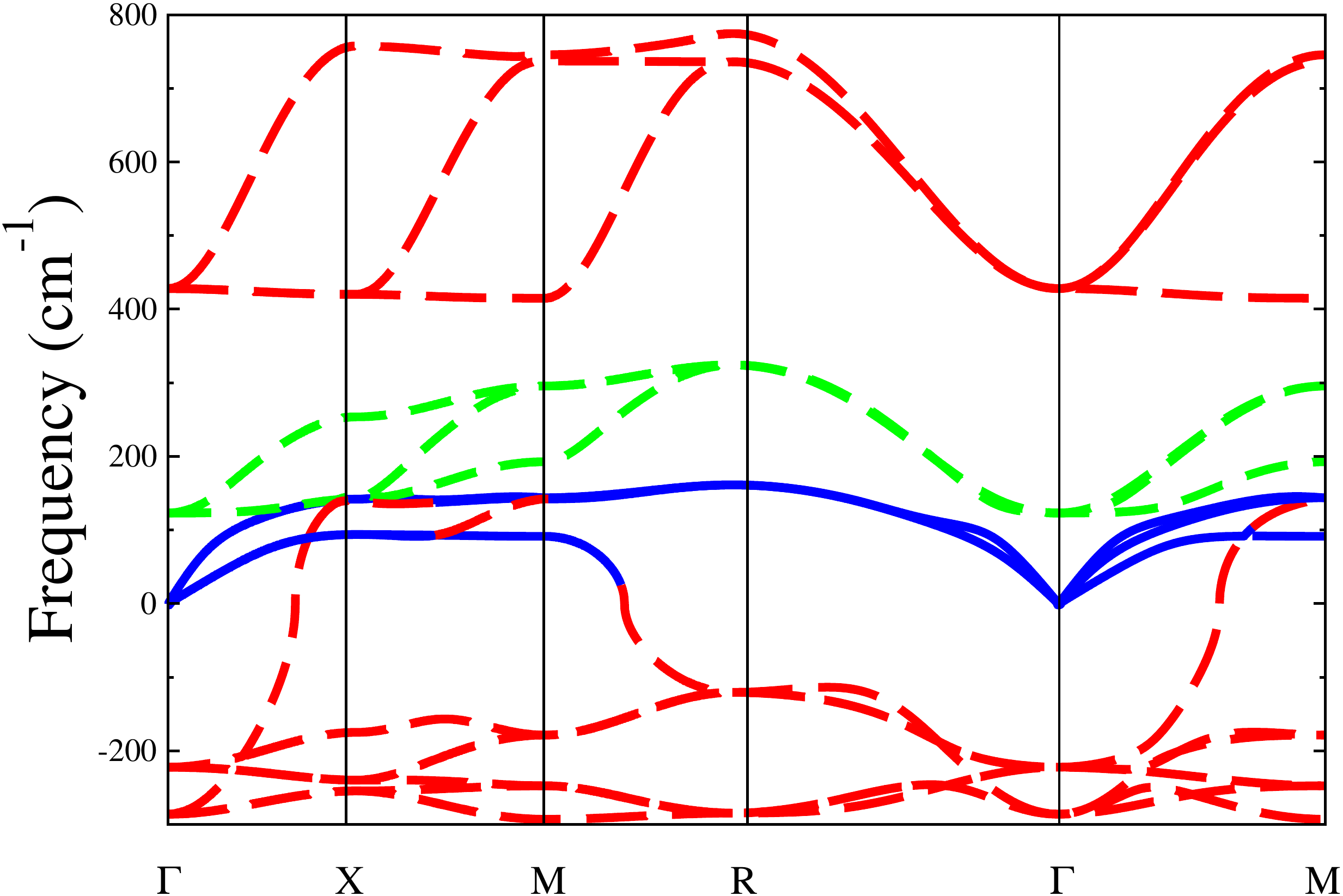}          
\caption
{(Color online) The phonon dispersion curve of SrPdO$_3$ with cubic (P$_{m\bar{3}m}$) space group at level of HSE06. The dark full (blue),
light-gray dashed (green) and dark-gray dashed (red) lines indicate a predominantly Sr, Pd, and O character of the dynamic
matrix eigenvectors associated with the phonon curves, respectively. The negative phonon branches are mostly associated
with O instabilities.
}
\label{fig:2}
\end{figure}

In cubic perovskite, the only free structural parameter is the lattice constant $a$. The relaxed
lattice constant $a$ is predicted to be 3.946 \AA ~at HSE06 level. Spin-polarized calculations shown that the cubic phase
of SrPdO$_3$ is a non-magnetic metal, in which the Pd$^{4+}$ cation adopts the LS configuration
($d^{\uparrow\downarrow\uparrow\downarrow\uparrow\downarrow}$). Based on the relaxed structure,
the HSE06 calculated phonon dispersion curve along the high-symmetry directions
$\Gamma$-X-M-R-$\Gamma$-M is shown in Fig.\ref{fig:2}.  As expected from the above considerations on the tolerance factor,
there are many unstable (negative phonon frequency) phonon branches, 
indicating that the cubic phase of SrPdO$_3$ is dynamically unstable at low temperature.
The unstable modes are predominantly associated with O atoms displacement (dashed thick (red) line in Fig.\ref{fig:2}):
The most negative mode at the zone center is the ferroelectric (FE) instability $\Gamma_4^-$ (we use $A$ site center setting in this paper),
which represents the opposite displacement of Pd ion and O octahedra as indicated by the arrows in Fig.\ref{fig:1}b,
whereas at $M$ and $R$ the phonon instabilities are associated with the rotation and tilting of oxygen octahedra (antiferrodistortive, AFD)
$M_3^+$ and $R_4^+$, respectively (schematically shown in Fig.\ref{fig:1} c and d).\cite{howard98}
These instabilities are very common in cubic perovskites and were very well studied.\cite{Ghosez99,amisi12,Machado11}
Having established the phonon instabilities of the ideal cubic phase we now present the outcomes of our structural search,
which is based on the stabilization of the negative phonon modes through suitable atomic displacements.

\subsection{Structural search and determination of the most stable phase}
\label{sec:st}
From the phonon analysis of the cubic phase, it was concluded that there are three important unstable
modes: $\Gamma_4^-$, $M_3^+$, and $R_4^+$. In order to stabilize these unstable modes, we have displaced
the atoms according to the force constant eigenvectors and created a pool of subgroup structures.
Each of the obtained subgroup structure was then subjected to a full structural relaxation (unit cell
shape, volume, and atom positions) within the specific subgroup symmetry, i.e. during the relaxation procedure the
subgroup symmetry was kept fixed. The comparison of the total energies of the fully relaxed subgroup structures
will identify the most stable structural variant.

Clearly, the stabilization of negative phonon modes by moving the atomic positions breaks the cubic symmetry
and leads to lower symmetry structures. The total distortion amplitudes of the fully relaxed subgroup structures
with respect to cubic phase (reference phase) were calculated by using the Bilbao Crystallographic Server,\cite{bilbao}
and the direction of the order parameter (which is identified by the space group symmetries of the parent and subgroup
structure in a phase transition) is obtained according to the group theoretical methods collected in the ISOTROPY Software Suite.\cite{isotropy}
The results are collected in Tab. \ref{tab:1} where, for each subgroup structure, we summarize the following
information: (i) the number of formula units (f.u.) per primitive cell ($Z$), (ii) the characteristic mode,
(iii) the direction of the order parameter ($Dir$), (iv) the total distortion amplitude ($Amp$), and finally (v) the total energy
of the fully relaxed subgroup structure with respect to the ideal cubic phase ($\Delta{E}$). The various structural variants included in
Tab. \ref{tab:1} are grouped according to the distinctive unstable modes. Beside including the main unstable modes
$\Gamma_4^-$, $M_3^+$, and $R_4^+$, we have also considered
the other zone boundary unstable mode, i.e. $X$ ($X_5^+$), a few second largest instabilities at high symmetry
points $\Gamma$ ($\Gamma_5^-$), $R$ ($R_5^+$), and $M$ ($M_5^+$), as well as suitable combinations of these modes as detailed below.

\begin{table}
\caption{Collection of the relative energies ($\Delta$E, meV/f.u.), unstable modes (Modes), number of formula unit (f.u.)
per primitive cell ($Z$), direction of order parameter ($Dir$), total distortion amplitude ($Amp$, \AA) of all subgroup structures
considered in the structural search and obtained by stabilizing the unstable modes of cubic SrPdO$_3$.
} \vspace{0.3cm}
\begin{ruledtabular}
\begin{tabular}{lcccccc}
Subgroup             &     $Z$     &      Modes         &    $Dir$      &     $Amp$       & $\Delta$E   \\
$P_{4mm}$ (99)       &     1       &   $\Gamma_4^-$     &    (a,0,0)    &    0.2570       &   -136      \\
$A_{mm2}$ (38)       &     1       &   $\Gamma_4^-$     &    (a,a,0)    &    0.2736       &   -150      \\
$R_{3m}$ (160)       &     1       &   $\Gamma_4^-$     &    (a,a,a)    &    0.2608       &   -150      \\
$P_m$ (6)            &     1       &   $\Gamma_4^-$     &    (a,b,0)    &    0.2637       &   -148     \\
$C_m$ (8)            &     1       &   $\Gamma_4^-$     &    (a,a,b)    &    0.2658       &   -160      \\
$P_{\bar{4}m2}$ (115)&     1       &   $\Gamma_5^-$     &    (a,0,0)    &    0.2950       &    -64      \\
$A_{mm2}$ (38)       &     1       &   $\Gamma_5^-$     &    (a,a,0)    &    0.4363       &   -144      \\
$R_{32}$ (155)       &     1       &   $\Gamma_5^-$     &    (a,a,a)    &    0.4088       &   -133      \\
$I_{4/mcm}$ (140)    &     2       &    $R_4^+$         &    (a,0,0)    &    0.9619       &   -439      \\
$I_{mma}$ (74)       &     2       &    $R_4^+$         &    (a,a,0)    &    0.9834       &   -522      \\
$R_{\bar{3}c}$ (167) &     2       &    $R_4^+$         &    (a,a,a)    &    1.0040       &   -512      \\
$C_{2/m}$ (12)       &     2       &    $R_4^+$         &    (a,b,0)    &    0.9890       &   -523      \\
$C_{2/c}$ (15)       &     2       &    $R_4^+$         &    (a,a,b)    &    0.9976       &   -517      \\
$I_{4/mmm}$ (139)    &     2       &    $R_5^+$         &    (a,0,0)    &    0.1903       &   -11       \\
$P_{4/mbm}$ (127)    &     2       &    $M_3^+$         &    (a,0,0)    &    0.9551       &   -405      \\
$I_{4/mmm}$ (139)    &     4       &    $M_3^+$         &    (a,a,0)    &    1.3074       &   -449      \\
$I_{m\bar{3}}$ (204) &     4       &    $M_3^+$         &    (a,a,a)    &    1.2389       &   -445      \\
$I_{mmm}$ (71)       &     4       &    $M_3^+$         &    (a,b,c)    &    1.2849       &   -442      \\
$P_{mma}$ (51)       &     2       &    $M_5^-$         &(a,0,0,0,0,0)  &    0.2459       &    -5       \\
$C_{mmm}$ (65)       &     2       &    $M_5^-$         &(a,a,0,0,0,0)  &    0.0025       &   -91       \\
$R_{3m}$ (160)       &     4       &    $M_5^-$         &(a,0,a,0,a,0)  &    0.3649       &   -118      \\
$P_{2/m}$ (10)       &     2       &    $M_5^-$         &(a,b,0,0,0,0)  &    0.0026       &   -63       \\
$C_{mme}$ (67)       &     2       &    $M_5^+$         &(a,a,0,0,0,0)  &    0.0126       &   -19       \\
$P_{mna}$ (53)       &     2       &    $M_5^+$         &(a,a,0,0,a,-a) &    0.0454       &   -5        \\
$C_{mcm}$ (63)       &     2       &      $X_5^+$       &(a,0,0,0,0,0)  &    0.3643       &   -53      \\
$P_{mma}$ (51)       &     2       &      $X_5^+$       &(a,a,0,0,0,0)  &    0.3898       &   -57       \\
$I_{ma2}$ (46)          &     2       &$R_4^+\oplus \Gamma_4^-$ &(a,a,0),(a,0,0)& 0.9789, 0.0161  &  -517   \\
$R_{3c}$ (161)         &     2       &$R_4^+\oplus \Gamma_4^-$ &(a,a,a),(a,a,a)& 0.9846, 0.0130  &   -513      \\
$P_{nma}$ (62)        &     4       &$R_4^+\oplus M_3^+$    &(a,a,0),(a,0,0)& 1.1814, 0.9727  &   -576      \\
$C_{mcm}$ (63)       &     4       &$R_4^+\oplus M_3^+$    &(a,0,0),(a,0,0)& 0.9850, 1.0374  &   -542      \\
$P_{4/mbm}$ (127)  &     4       &$R_4^+\oplus M_3^+$    &(a,0,0),(a,0,0)& 0.6874, 0.7218  &   -224      \\
\end{tabular}
\end{ruledtabular}
\label{tab:1}
\end{table}

Condensing the largest unstable polar mode $\Gamma_4^-$ in different directions leads to five polar subgroup
structures: $P_{4mm}$, $A_{mm2}$, $R_{3m}$, $P_m$, and $C_m$, each characterized
by a distinctive direction of the order parameter, see Tab. \ref{tab:1}.
Both the energy gain and distortion amplitude
of these structures with respect to the reference cubic phase are very similar:
$\Delta$E $\approx$150 meV/f.u. and $Amp$ $\approx$ 0.26 \AA. This situation does not change much in terms of total
energy stability by considering the stabilization of the non-polar mode at the zone center $\Gamma_5^-$,
which guides the transition towards three low symmetry structures ($P_{\bar{4}m2}$, $A_{mm2}$, and $R_{32}$).
Much lower energy structures were obtained by condensing the tilting mode $R_4^+$ and rotation mode $M_3^+$.
The $M_3^+$ and $R_4^+$ derived subgroup structures
($M_3^+$ $\rightarrow$ $P_{4/mbm}$, $I_{4/mmm}$, $I_{m\bar{3}}$, and $I_{mmm}$;
$R_4^+$ $\rightarrow$ $I_{4/mcm}$, $I_{mma}$, $R_{\bar{3}c}$ , $C_{2/m}$, and $C_{2/c}$)
are strongly distorted ($Amp$ is in the range 0.95 $\sim$1.31 \AA) and show energy gain of about 400 ($M_3^+$) and 500 ($R_4^+$) 
meV/f.u..
It is worth pointing out that the distortion amplitude of the $R_4^+$ derived structures are
not larger than that of $M_3^+$, indicating that the stiffness coefficients
(defined as $k=(1/2)(\partial^2E/\partial^2Q)$, where $Q$ is the amplitude of mode) of the $R_4^+$ distortion is
significantly larger than that of $M_3^+$ mode. It is obvious that, on the other hand,
the stabilization of the second largest unstable phonon modes $R_5^+$, $M_5^+$, and $X_5^+$ leads to much less favorable structures.

\begin{figure}
\includegraphics[clip=,width=0.50\textwidth]{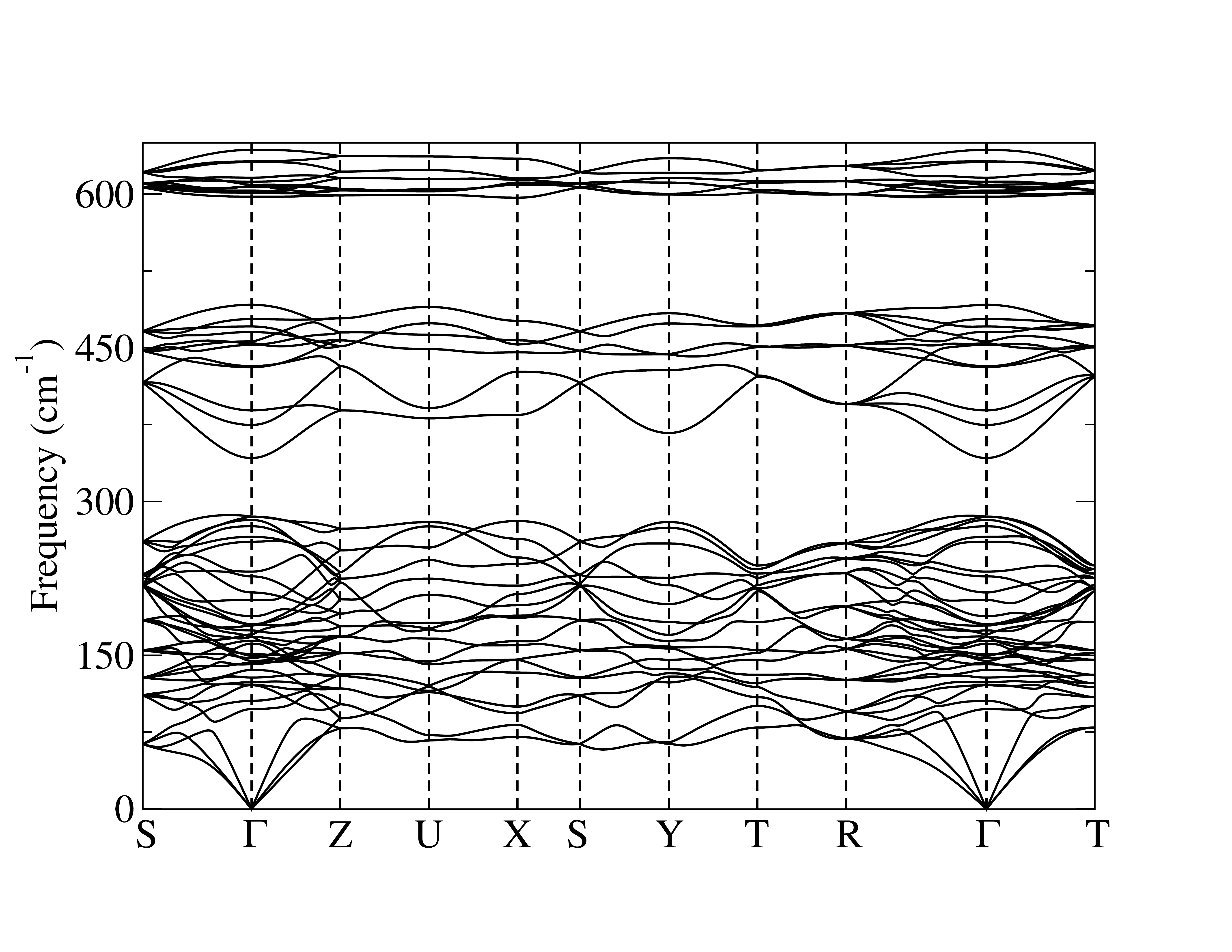}
\caption
{Phonon dispersion curves of the $P_{nma}$ phase at PBE level.
}
\label{fig:3}
\end{figure}

The fact that the amount of energy gained by condensing the single unstable modes
$M_3^+$ and $R_4^+$ are much larger than that of $\Gamma_4^-$ and of the other unstable modes indicates that in SrPdO$_3$
AFD instabilities are much stronger than FE ones. Therefore AFD instabilities should be considered as primary order
parameters, which is similar with other 4$d$ transition metal oxides such as NaNbO$_3$ and SrZrO$_3$.\cite{Machado11,amisi12}
Thus, in order to find possibly more stable structures we have considered the combination of $R_4^+$ and $M_3^+$ distortions
at the same time. Three possible combinations of AFD with different order parameter directions
were studied, as shown in Table \ref{tab:1}: $P_{nma}$, $P_{4/mbm}$, and $C_{mcm}$.
We found, that the most stable structure is $P_{nma}$ (No. 62), which is the combination of $R_4^+$ in (a,a,0) direction
and $M_3^+$ in (a,0,0) direction, and characterized by a energy gain of 576 meV/f.u.. \cite{howard98}
The amplitudes of $R_4^+$ and $M_3^+$ modes are almost identical to the corresponding single-mode frozen cases
indicating that these two modes are compatible with each other at the lower symmetry phase $P_{nma}$.
Moreover, it is found that the energy gain of the combined $R_4^+\oplus M_3^+$ distortion is larger than the corresponding
single-mode frozen cases ($R_4^+$ and $M_3^+$).

In order to check the dynamical stability of the $P_{nma}$ phase, we have calculated its phonon dispersion at PBE level
(cross-checked by LDA and PBEsol) and did not find any unstable mode, as shown in Fig. \ref{fig:3}. 
As a matter of fact, $P_{nma}$ is the most common space group of TM oxide perovskite structures
because it takes into account the flexibility of $B$O$_6$ oxygen octahedra network throughout collective rotations and tiltings
to optimize the coordinate environment of the cation $A$.\cite{woodward97,Lufaso01}
Interestingly, another perovskite structure $C_{mcm}$ has the second lowest energy among all subgroup structures.
The $C_{mcm}$ space group originates again from a combination of the $M_3^+$ and $R_4^+$ modes, but with different
order parameter directions. This further confirms that $R_4^+$ and $M_3^+$ are the most important unstable factors
in the high symmetry phase. Therefore, $P_{nma}$ could be identified as the ground state structure of SrPdO$_3$.

On the basis of these structural analysis, we can exclude the possibility of
antiferroelectricity in SrPdO$_3$. Although the antipolar mode $X_5^+$ is unstable and connects with the 
polar mode $\Gamma_4^-$ by an unstable branch, as shown in Fig. \ref{fig:2}, the energy gained by condensing $X_5^+$ ($\Delta E$ $\sim$ 60 meV/f.u.) 
and $\Gamma_4^-$ ($\Delta E$ $\sim$ 150 meV/f.u.) are much smaller than that of AFD ($\Delta{E}$ $\sim$ 500 meV/f.u.).
Thus, it is apparent that SrPdO$_3$ does not fully satisfy the criterion of antiferroelectrity as 
$X_5^+$ is too weak to compete with the AFD instability.\cite{rabe13}

\begin{figure}
\includegraphics[clip,width=0.49\textwidth]{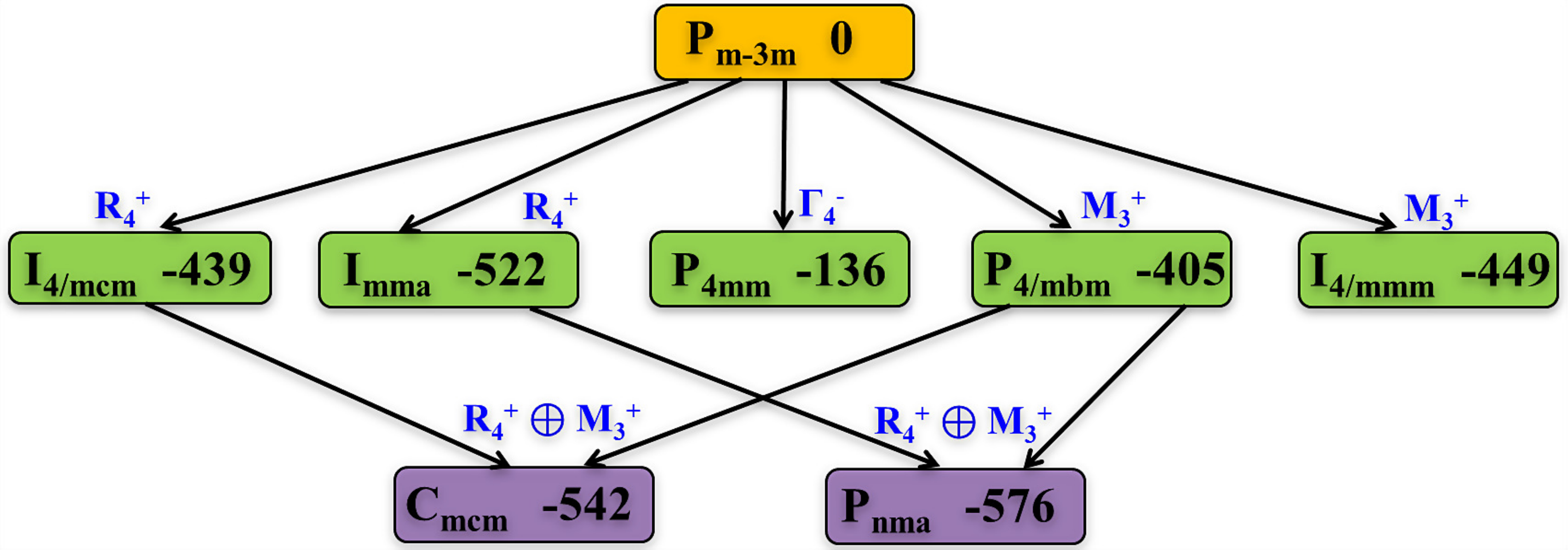}
\caption
{(Color online) Group and subgroup tree of high symmetry structure ($P_{m\bar{3}m}$)
and low-lying structures ($P_{nma}$ and $C_{mcm}$). The energy gain (meV/f.u.), by
condensing unstable modes (blue color), with respect to cubic phase is also
indicated using black number.
}
\label{fig:4}
\end{figure}

The most important group and subgroup connections associated with the main distortions
analyzed in this study are summarized in Fig. \ref{fig:4}. Starting from the undistorted cubic structure $P_{m\bar{3}m}$
the stabilization of the single modes $R_4^+$, $\Gamma_4^-$, and $M_3^+$ leads to a subset of subgroup structures
with different relative energies. The combination of $R_4^+$ and $M_3^+$ AFD modes guides the
transition towards the two low-lying structures: the $P_{nma}$ phase, ubiquitous in the perovskite family, and the
meta-stable structure $C_{mcm}$, which was observed in NaNbO$_3$ and proposed as intermediate phase of 
SrZrO$_3$.\cite{darlington99,amisi12}

The optimized structural parameters of $P_{nma}$ SrPdO$_3$ are tabulated in Table \ref{pnma}. Besides the HSE06 data, 
the results based on PBE and LDA are also shown for comparison. As compared with HSE06,  PBE overestimates all lattice 
constants, especially $a$, whereas LDA underestimates lattice constants $b$ and $c$ but overestimates $a$.

\begin{table}
\caption{The fully relaxed structural parameters (lattice constants $a$, $b$, and $c$, Volume $V_0$, and Wyckoff positions)
of ground state SrPdO$_3$ (space group  $P_{nma}$, No. 62) as obtained by HSE, PBE, and LDA.}
\vspace{0.3cm}
\begin{ruledtabular}
\begin{tabular}{ccccccc}
           &  Wycyoff site  &       x        &     y            &        z         \\
\multicolumn{5}{c}{HSE}                                                        \\
a=5.4933 \AA     &b=7.8124 \AA   &  c=5.5233 \AA& $V_0$ = 237 &    \\
Sr                &     4c                & 0.45925  & 0.25000   &   0.49305  \\
Pd               &     4a                & 0.00000  & 0.00000   &   0.00000  \\
O$_1$        &     4c                & 0.00464  & 0.25000   &   0.07464  \\
O$_2$        &     8d                & 0.70647  & 0.03753   &   0.79362  \\
\multicolumn{5}{c}{PBE}                                                             \\
a=5.6992 \AA     &b=7.9835 \AA   &  c=5.6402 \AA& $V_0$ = 257    &    \\
Sr               &     4c                & 0.45162  &  0.25000  &   0.49016  \\
Pd              &     4a                & 0.00000  &  0.00000  &   0.00000  \\
O$_1$       &     4c                & 0.01697  &  0.25000  &   0.08190  \\
O$_2$       &     8d                & 0.70456  &  0.04086  &   0.7973    \\
\multicolumn{5}{c}{LDA}                                                             \\
a=5.5218 \AA     &b=7.7942 \AA    &  c=5.5067  \AA&   $V_0$ = 237  &    \\
Sr              &     4c                & 0.45307   &  0.25000  &  0.49072  \\
Pd             &     4a                & 0.00000  &  0.00000  &   0.00000  \\
O$_1$      &     4c                & 0.00952  &  0.25000  &   0.07982  \\
O$_2$      &     8d                & 0.70471  &  0.03983  &   0.79627  \\
\end{tabular}
\end{ruledtabular}
\label{pnma}
\end{table}

We conclude the discussion of the structural search by comparing the structural distortions in SrPdO$_3$ with
those of other 4$d$ perovskites based on TM atoms which precede Pd in the periodic table: SrTcO$_3$, SrRuO$_3$, and SrRhO$_3$.
These compounds also crystallize with the orthorhombic $P_{nma}$ space group at low temperature.\cite{efrain,bushmeleva,jones,yamaura}
The most important structural data are collected in Tab.\ref{tab:2} as a function of the occupation of the
4$d$ shell going from SrTcO$_3$ (d$^3$) to SrPdO$_3$ (d$^6$).
For all these 4$d$ perovskites the tolerance factor is very similar, $\sim$ 0.9.
Owing to the full occupation of $t_{2g}$ orbitals (low-spin configuration), SrPdO$_3$ has a smaller volume, no Jahn-Teller (JT)
distortions ($Q_2$=0.004 and $Q_3$=-0.004) and the largest GaFeO$_3$-type (GFO) distortions.

\begin{table}
\caption{Structural parameters of SrTcO$_3$ (HSE data, Ref.\onlinecite{franchini11}), SrRuO$_3$ (Experiment, Ref.\onlinecite{jones}),
SrRhO$_3$ (Experiment, Ref.\onlinecite{yamaura}) and SrPdO$_3$ (HSE06, this study) within the $P_{nma}$ space group. The
Jahn-Teller $Q_2$ and $Q_3$ distortions are defined in accordance with the formula given in
Ref. \onlinecite{franchini11}:  $Q_2=2(l-s)/\sqrt(2)$ and $Q_3=2(2m-l-s)/\sqrt(6)$  where $l$, $s$, and $m$ refer to
long (l) and short (s) TM-O$_2$ in-plane distances and medium (m) TM-O$_1$ vertical ones.}
\vspace{0.3cm}
\begin{ruledtabular}
\begin{tabular}{ccccccc}
                        & SrTcO$_3$ $^a$  & SrRuO$_3$ $^b$& SrRhO$_3$ $^c$&   SrPdO$_3$   \\
4$d$ occupation         &    $d^3$        &    $d^4$      &    $d^5$      &    $d^6$      \\
TM Ionic radii (pm)$^d$ &    64.5         &     62.0      &     60.0      &    61.5       \\
tolerance factor        &    0.8921       &    0.9031     &    0.9122     &    0.9054     \\
Volume                  &    242.74       &    241.52     &    242.18     &    237.04     \\
$a$                     &    5.543        &    5.530      &    5.539      &    5.493      \\
$b$                     &    7.854        &    7.845      &    7.854      &    7.812      \\
$c$                     &    5.576        &    5.567      &    5.567      &    5.523      \\
TM-O$_1$                &    1.990        &    1.981      &    1.990      &    1.996      \\
TM-O$_{2}^l$            &    2.015        &    1.988      &    2.040      &    2.001      \\
TM-O$_{2}^s$            &    1.942        &    1.984      &    1.970      &    1.996      \\
$Q_2$                   &    0.103        &    0.003      &    0.050      &    0.004      \\
$Q_3$                   &    0.018        &   -0.008      &   -0.025      &   -0.004      \\
TM-O$_1$-TM             &   168.76        &    162.79     &    161.10     &    156.12     \\
TM-O$_2$-TM             &   161.57        &    162.75     &    156.52     &    154.06     \\
\end{tabular}
\end{ruledtabular}
\label{tab:2}
\begin{flushleft}
$^a$Ref: \cite{franchini11}
$^b$Ref: \cite{jones}
$^c$Ref: \cite{yamaura}
$^d$Ref: \cite{shannon}
\end{flushleft}
\end{table}

\subsection{Ground state electronic structure}
\label{sec:el}
 
As mentioned previously, spin-polarized calculations show that all the structures considered in this paper
are non-magnetic (low spin state, $t_{2g}^{\uparrow\downarrow\uparrow\downarrow\uparrow\downarrow}e_g^0$).
In this section we present the electronic properties of the ground state non-magnetic P$_{nma}$ phase of 
SrPdO$_3$ based on HSE and many-body GW techniques. 

\begin{figure}
\includegraphics[clip,width=0.45\textwidth]{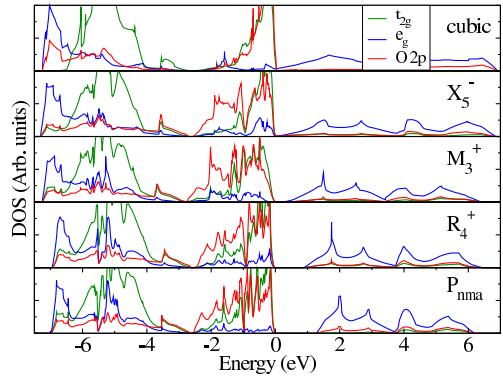}
\caption
{(Color online) The density of state of SrPdO$_3$ without distortion (cubic), with distinct $M_3^+$, $R_4^+$, and
$X_5^+$ distortions, and within P$_{nma}$ symmetry at HSE level for $\alpha=0.25$.}
\label{fig:5}
\end{figure}

\begin{figure}
\includegraphics[clip,width=0.45\textwidth]{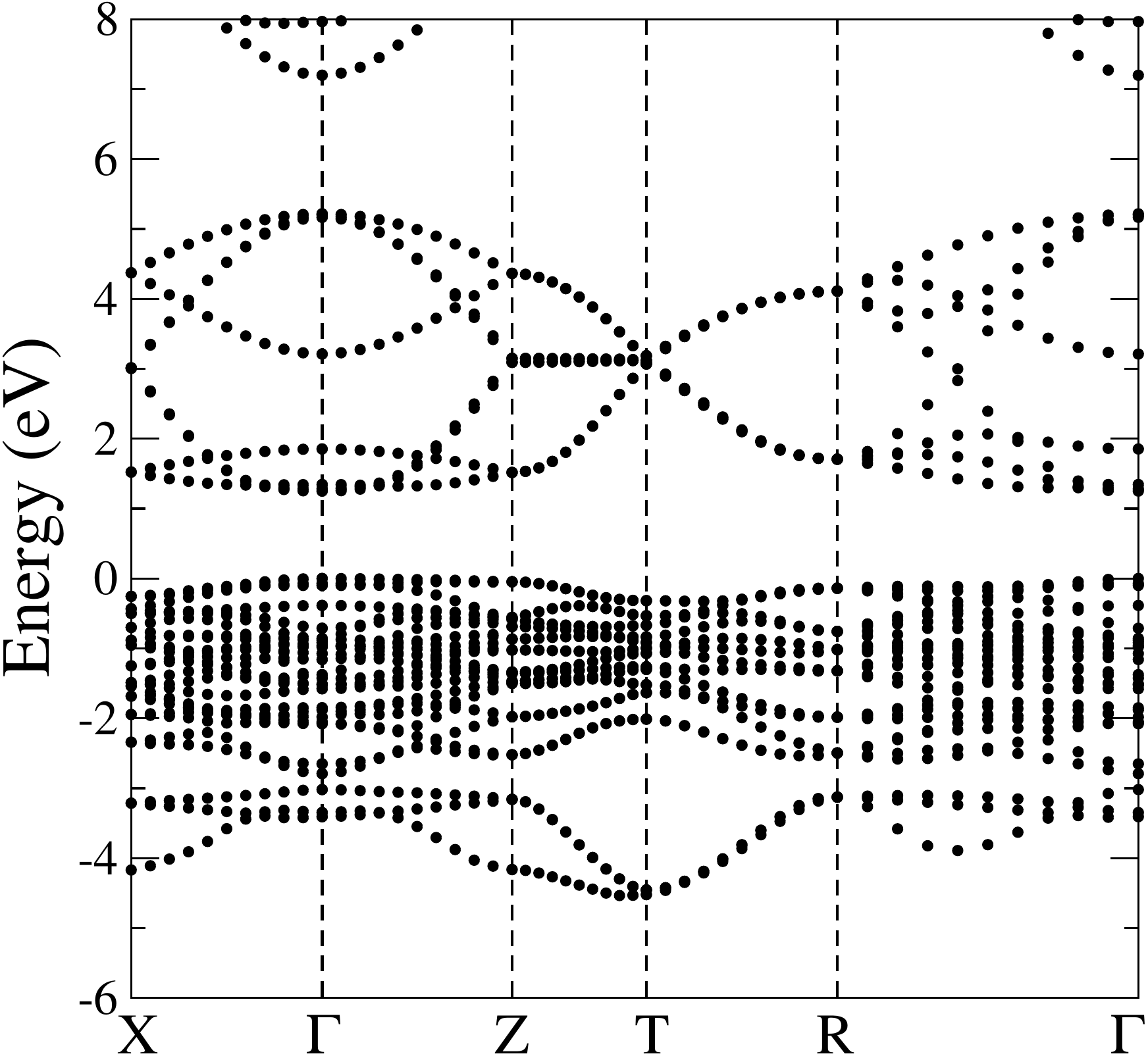}
\caption
{Band structure P$_{nma}$-structured SrPdO$_3$ as computed by HSE for $\alpha=0.25$.}
\label{fig:6}
\end{figure}

At HSE level, using the 'standard' value of the mixing parameter $\alpha$=0.25 we find that P$_{nma}$ SrPdO$_3$
is a direct band gap semiconductor with a band gap of 1.27 eV. the gap is opened between the fully occupied $t_{2g}$ shell 
(hybridized with the O $p$ states) and the unoccupied $e_g$ manifold laying on the bottom of the conduction band
as illustrated in Figs. \ref{fig:5} and \ref{fig:6}. 
The electronic dispersion in Fig. \ref{fig:6} looks similar to the one computed for LaCoO$_3$,\cite{he12,nohara,Hsu10,knizek} and 
is characterized by a valence/conduction band region dominated by $t_{2g}$/$e_g$ orbitals significantly hybridized with O-$p$ states, in 
agreement with a formal LS state. The major difference is the bandwidth of the bottom of the conduction band which is much larger in 
SrPdO$_3$ (4.0 eV) as compared to LaCoO$_3$ (3 eV), which reflects the more delocalized nature of 4$d$ electrons.\cite{franchini11}

In order to clarify the role of the predominant structural distortions discussed  
in the previous section on the electronic properties we compare in Fig. \ref{fig:5} the density of states (DOS) 
of the ideal cubic and P$_{nma}$ cases with those corresponding to the structures containing only one single  
mode ($M_3^+$, $R_4^+$, and $X_5^+$) at standard (i.e. $\alpha$=0.25) HSE06 level. The cubic phase turns out to be metallic, with a very wide $e_g$
unoccupied band crossing the Fermi level. The $R_4^+$ induces a substantial narrowing of the $e_g$ bands
which leads to the opening of a rather large band gap of about 0.9 eV. This is mainly due to the cooperative 
rotation and tilting of the oxygen octahedra (see Fig.\ref{fig:1}) which significantly reduces the Pd $e_g$-O 2$p$ hybridization.
Conversely, both the $M_3^+$ (in-plane oxygens rotation) and the $X_5^+$ (Sr anti-polar displacements) modes are not sufficiently 
effective in weakening the Pd-O hybridization and therefore are unable to open a sizable gap.

\begin{figure}
\includegraphics[clip,width=0.45\textwidth]{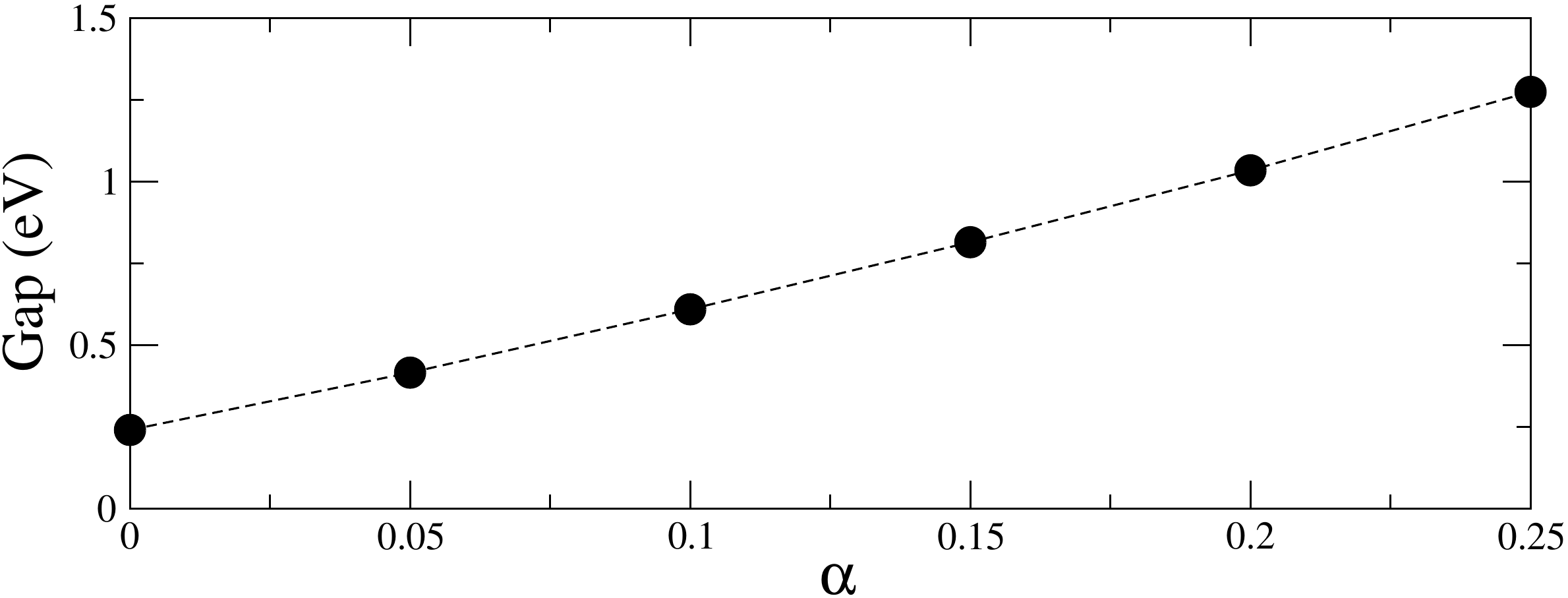}
\caption
{Dependence of the HSE band gap as a function of the mixing parameter $\alpha$.}
\label{fig:7}
\end{figure}

We now address the issue of band gap prediction.
The value of the band gap at hybrid functional level depends on the value of the mixing factor
$\alpha$, which determines the amount of exact Hartree-Fock exchange incorporated in the HSE functional.
More specifically, there is an almost linear dependence between the band gap and $\alpha$, as shown for SrPdO$_3$ in 
Fig. \ref{fig:7}: we found that PBE (equivalent to $\alpha=0$) is sufficient to open a small gap of 0.24 eV, and
by increasing $\alpha$ the gap growths linearly up to 1.27 eV for $\alpha$=0.25 (notice that the choice of $\alpha$ is much 
less crucial for what concern the structural properties which are well described (i.e. within 1 \%) for any reasonable 
value of $\alpha$). We have discussed this issue in detail in Ref. \onlinecite{he12} where we have used a 
semiempirical rationale for the determination of the optimum mixing factor for HSE in 3$d$ transition metal perovskites 
which requires the knowledge of the dielectric constant ${\epsilon_{\infty}}$.\cite{Alkauskas10} 
Unfortunately, the ${\epsilon_{\infty}}$ of SrPdO$_3$ is unknown and this hampers a precise estimation of the optimum 
$\alpha$ for SrPdO$_3$, as the {\em ab initio} calculation of the dielectric properties
is by far not trivial.\cite{Alkauskas10, Paier08} 

The most accurate way to overtake this limitation is to go beyond generalized Kohn-Sham (gKS) orbitals such as HSE and rely on 
more sophisticated and computationally much more demanding many-body techniques such as GW.\cite{hedin}
The performance and the predictive power of GW-based approaches critically depends on the treatment of the 
dynamically screened Coulomb potential W and, to a lesser extent, on the self-consistency in G. 
It has been shown that the results obtained within the widely used single-shot G$_0$W$_0$ and partially 
self-consistent GW$_0$ approximations depend significantly on the starting point for the solution of the quasiparticle 
equation.\cite{Fuchs07,Paier08}
In general, gKS functionals are considered to be good starting points only if electron-hole interactions are incorporated. 
This is due to the fact that GW$_0$@HSE within the so called random phase approximation (RPA, i.e. excluding excitonic
effects in the calculation of the screened Coulomb interaction) leads to a significant underestimation of the static screening 
resulting in too large band gap.\cite{Paier08}
Conversely, considering that LDA and PBE functionals usually provide reasonable screening properties, the G$W_0$@PBE 
usually deliver band gaps in good agreement with experiment. \cite{Paier08}
Following these arguments here we compare the results obtained from two different procedures:
(i) GW$_0$@PBE (partially self consistent GW on top of PBE orbitals within the RPA approximation), 
and (ii) GW$_0$-TCTC@HSE (partially self-consistent GW on top of HSE orbitals plus inclusion of excitonic effects
via the TCTC kernel\cite{Bruneval05}). Clearly, the optimum approach would require a self-consistent computation of W, 
which would render the corresponding GW results free of any dependence on the starting orbitals. Unfortunately, the enormous 
computational load associated with this scheme prevents its applicability to the present case, mostly due to large size of the 
unit cell which contains 20 atoms. To date, the only application of fully self-consistent quasi-particle GW plus vertex corrections 
for non-cubic perovskite is the case of BaBiO$_3$, whose unit cell contains {\em only} 10 atoms.\cite{Franchini10}    

The results on the band gap are collected in Tab.\ref{tab:3}. The application of GW always leads to an increase of the band gap with 
respect to the corresponding starting PBE and HSE value (first row of Tab. \ref{tab:3}. The converged GW$_0$@PBE value is 0.92 eV, whereas
the  GW$_0$-TCTC@HSE estimations ranges between 1.03 eV to 1.31 eV, depending on the value of $\alpha$. As shown in the 
last column of Tab. \ref{tab:3} the relative deviations among the various schemes are intimately connected with
the computed values of the dielectric constant $\epsilon_{\infty}$, which progressively decreases as a function
of $\alpha$. By averaging over the different starting points our GW$_0$ data suggest that the band gap of SrPdO$_3$ is 1.1 $\pm$ 0.2 eV.

\begin{table*}
\caption{Band gap of SrPdO$_3$ calculated by using HSE with different $\alpha$ ($\alpha$=0 refers to PBE calculations), 
GW$_0$ and GW$_0$-TCTC. The value of $\epsilon_{\infty}$ computed within GW$_0$-TCTC is also shown.}
\vspace{0.3cm}
\begin{ruledtabular}
\begin{tabular}{lcccccc}
 $\alpha$                  &  0.00    &  0.05  &   0.10   &  0.15  &  0.20  &    0.25   \\  \hline
 HSE($\alpha$)             &  0.24    &  0.41  &   0.61   &  0.81  &  1.03  &    1.27   \\  
 GW$_0$@HSE($\alpha$)      &  0.92    &  1.08  &   --     &  1.31  &   --   &    1.43   \\
 GW$_0$-TCTC@HSE($\alpha$) &  --      &  1.03  &   --     &  1.21  &   --   &    1.31   \\       
 $\epsilon_{\infty}$       & 16.6     & 13.7   &   --     & 10.9 &   --   &    9.5                   
\end{tabular}
\end{ruledtabular}
\label{tab:3}
\end{table*}

\section{SUMMARY AND CONCLUSIONS}

Summing up, in this study we have performed an accurate structural determination of the novel
perovskite SrPdO$_3$, based on an appropriate stabilization of the unstable phonon modes found in the ideal perovskite phase
by applying symmetry analysis and group theory considerations.
We found that the combination of AFD displacements associated with the $R_4^+$ and $M_3^+$
modes drives the transition from the unstable cubic phase to the orthorhombic $P_{nma}$ ground state,
characterized by a large tilting of the BO$_6$ oxygen octahedra as reflected by the significant GFO-like distortions.
This is in line with the structural properties of the 4$d$ perovskite series Sr$R$O$_3$, with $R$=Tc, Ru, and Rh.
The ground state electronic structure was studied by using partially self-consistent GW on top of both PBE and HSE 
orbitals, and including electron-hole interaction in the framework of the TCTC scheme. Our results indicate that 
low temperature SrPdO$_3$ is a low-spin state insulator with a band gap of $\approx$ 1.1 eV. The electronic transition from 
the metallic ideal cubic phase to the insulating $P_{nma}$ one is mostly driven by the structural distortions associated 
with the $R_4^+$ mode which strongly attenuates the O$_p$-Pd$_{e_g}$ hybridization.
The experimentally observed high-temperature antiferromagnetic behavior might originate from spin state excitations,
possibly subjected to spin-state crossovers, in similarity with the case of isoelectronic LaCoO$_3$.
We hope that our study will spur further research aiming to clarify the spin structure of this novel perovskite,
for instance by investigating its nuclear hyperfine interaction, to provide an experimental measurement of the band gap
in support to our predictions, and, from an application point of view, will help the understanding and optimization of 
the reported high catalytic activity, and the designing of novel materials with tunable functionalities based on the 
$P_{nma}$ building block.\cite{Rondinelli.AM.2012,Mulder.AFM.2013}

\section{ACKNOWLEDGMENTS}
Research in Vienna was sponsored by the FP7 European Community grant ATHENA.
All calculations were performed at the Vienna Scientific Cluster (VSC).


\begin{thebibliography}{99}
\bibitem{Vrejoiu08}  I. Vrejoiu, M. Alexe, D. Hesse, and U. G\"osele,
                     Adv. Func. Mater. {\bf 18}, 3892 (2008).

\bibitem{maeno} Y. Maeno, H. Hashimoto, K. Yoshida, S. Nishizaki, T. Fujita, J. G. Bednorz,
                and F. Lichtenberg, Nature (London) {\bf 372}, 532 (1995).
              
\bibitem{Habermeier07} H. U. Habermeier, Mater. Today {\bf 10}, 34 (2007)


\bibitem{kimura} T. Kimura, Annu. Rev. Mater. Res. {\bf 37}, 387 (2007).

\bibitem{Ramesh07} R. Ramesh, N. A. Spaldin, Nat. Mater. {\bf 6}, 21 (2007).

\bibitem{Stroppa12}
A. Stroppa, P. Barone, P. Jain, J. M. Perez-Mato, S. Picozzi,
Adv. Mater. {\bf 25} 2284 (2013).

\bibitem{adler} S. B. Adler, Chem. Rev. {\bf 104}, 4791 (2004).


\bibitem{Suntivich11} J. Suntivich, H. A. Gasteiger, N. Yabuuchi, H. Nakanishi, J. B. Goodenough
                                       and Y. Shao-Horn, Nature Chem. {\bf 3} 546 (2011).

\bibitem{Misono05} M. Misono, Cat. Today {\bf 100}, 95 (2005).

\bibitem{Ohta07} H. Ohta, Mater. Today {\bf 10}, 44 (2007).

\bibitem{Lee03} Y. S. Lee, J. S. Lee, T. W. Noh, D. Y. Byum, K. Soo Yoo, K. Yamaura, and E. Takayama-Muromachi,
                Phys. Rev. B {\bf 67}, 113101 (2003).             

\bibitem{allen} P. B. Allen, H. Berger, O. Chauvet, L. Forro, T. Jarlborg, A. Junod, B. Revaz, and G. Santi,
                Phys. Rev. B {\bf 53}, 4393 (1996).

\bibitem{kostic} P. Kostic, Y. Okada, N. C. Collins, and Z. Schlesinger,
                 Phys. Rev. Lett. {\bf 81}, 2498 (1998).
 
\bibitem{Rondinelli08} J. M. Rondinelli, N.  M. Caffrey, S. Sanvito, and N. A. Spaldin
                 Phys. Rev. B {\bf 78}, 155107 (2008).

\bibitem{nagai} I. Nagai, N. Shirakaw, S. Ikeda, R. Iwasaki, H. Nishimura, and M. Kosaka,
                Appl. Phys. Lett. {\bf 87}, 024105 (2005).                 

\bibitem{efrain} E. E. Rodriguez, F. Poineau, A. Llobet, B. J. Kennedy, M. Avdeev, G. J. Thorogood,
                 M. L. Carter, R. Seshadri, D. J. Singh, and A. K. Cheetham,
                  Phys. Rev. Lett. {\bf 106}, 067201 (2011).

\bibitem{franchini11} C. Franchini, T. Archer, J. He, X. Chen, A. Filippetti, and S. Sanvito,
                    Phys. Rev. B {\bf 83}, 220402(R) (2011).
 
\bibitem{Middey12} S. Middey, A. Kumar Nandy, S. K. Pandey, P. Mahadevan, and D. D. Sarma,
                   Phys. Rev. B {\bf 86}, 104406, (2012).

\bibitem{Mravlje12} J. Mravlje, M. Aichhorn, and A. Georges, Phys. Rev. Lett. {\bf 108}, 197202 (2012). 


\bibitem{Tanaka02} Y. Nishihata,  J. Mizuki, T. Akao, H. Tanaka, M. Uenishi, M. Kimura, T. Okamoto, and N. Hamada, Nature {\bf 418}, 164 (2002).

\bibitem{Sun10} C. Sun, R. Hui, and J. Roller, J. Solid State Electrochem {\bf 14}, 1125 (2010).

\bibitem{yeung99} J. A. Yeung, K. Chen, A. T. Bell, and E. Iglesia,
                  J. Catal. {\bf 188}, 132 (1999).

\bibitem{meng11} L. Meng, A. P. Jia, J. Q. Lu, L. F. Luo, W. X. Huang, and M. F. Luo,
                J. Phys. Chem. C {\bf 115}, 19789 (2011).                        
                 
\bibitem{galal} A. Galal, N. F. Atta, S. A. Darwish, A. A. Fatah, S. M. Ali,
                 J. Power Sources, {\bf 195}, 3806 (2010).

\bibitem{Goodenough01} J.B. Goodenough, {\em Localized to Itinerant Electronic Transition
in Perovskite Oxides}, (Springer, New York, 20011), and references therein.

\bibitem{Rao04}  C. N. R. Rao, Md. M. Seikh, and C. Narayana, Top. Curr. Chem {\bf 234}, 1 (2004), and references therein.

\bibitem{saitoh} T. Saitoh, T. Mizokawa, A. Fujimori, M. Abbate, Y. Takeda, and M. Takano,
                 Phys. Rev. B {\bf 55}, 4257 (1997).

\bibitem{tokura} Y. Tokura, Y. Okimoto, S. Yamaguchi, and H. Taniguchi,
                 Phys. Rev. B {\bf 58}, R1699 (1998).                 

\bibitem{Raccah67} P. M. Raccah and J. B. Goodenough, Phys. Rev. {\bf 155}, 932 (1967).

\bibitem{Hsu10} H. Hsu, P. Blaha, R. M. Wentzcovitch, and C. Leighton
                Phys. Rev. B {\bf  82}, 100406(R) (2010), and references therein.               

\bibitem{parlinski00} K. Parlinski and Y. Kawazoe, Eur. Phys. J. B {\bf 16}, 49 (2000).

\bibitem{lebedev} A.I. Lebedev, Phys. Solid State {\bf 51} , 341  (2009).      

\bibitem{gk1} G. Kresse and J. Hafner,  Phys. Rev. B {\bf 48}, 13115 (1993).

\bibitem{gk2} G. Kresse and J. Furthm{\"u}ller,  Comput. Mater Sci. {\bf 6}, 15 (1996).

\bibitem{blochl} P. E. Bl\"ochl,
              Phys. Rev. B {\bf 50}, 17953 (1994).

\bibitem{gk4} G. Kresse and D. Joubert, Phys. Rev. B {\bf 59}, 1758 (1999).


\bibitem{heyd} J. Heyd, G. E. Scuseria, and M. Ernzerhof,
                  J. Chem. Phys. {\bf 118}, 8207 (2003); {\bf 124}, 219906 (E) (2006).

\bibitem{phonon1} K. Parlinski, Z. Q. Li, and Y. Kawazoe, Phys. Rev. Lett. {\bf 78}, 4063 (1997).

\bibitem{phonopy} A. Togo, F. Oba, and I. Tanaka, Phys. Rev. B {\bf 78}, 134106 (2008).

\bibitem{PBE}   John P. Perdew, Kieron Burke, and Matthias Ernzerhof,
            Phys. Rev. Lett. {\bf 77},  3865 ((1996).    

\bibitem{LDA} D. M. Ceperley and B. J. Alder, Phys. Rev. Lett. {\bf 45}, 566 (1980).
     
\bibitem{PBEsol} J. P. Perdew, A. Ruzsinszky, G. Csonka, O. A. Vydrov, G. E. Scuseria, L. A. Constantin, X. Zhou and K. Burke,
                           Phys. Rev. Lett. {\bf 100}, 136406 (2008).
                           
\bibitem{hedin} L. Hedin, Phys. Rev. {\bf 139}, A796 (1965).                               

\bibitem{Shishkin1}
M. Shishkin and G. Kresse, Phys. Rev. B {\bf 74}, 035101 (2006).

\bibitem{Shishkin2}
M. Shishkin and G. Kresse, Phys. Rev. B {\bf 75}, 235102 (2007).

\bibitem{Fuchs07}
F. Fuchs, J. Furthm\"uller, F. Bechstedt, M. Shishkin and G. Kresse,
Phys. Rev. B {\bf 76}, 115109 (2007).

\bibitem{Paier08}
Joachim Paier, Martijn Marsman, and Georg Kresse,
Phys. Rev. B {\bf 78}, 121201(R) (2008).

\bibitem{Franchini10}
C. Franchini, A. Sanna, M. Marsman, and G. Kresse,
Phys. Rev. B {\bf 81},  085213 (2010).

\bibitem{Bruneval05}
F. Bruneval, F. Sottile, V. Olevano, R. Del Sole, and L. Reining, Phys. Rev. Lett.,
{\bf 94}, 186402 (2005).

\bibitem{he12} J. He and C. Franchini,
               Phys. Rev. B {\bf 86}, 235117 (2012).   
               
\bibitem{Goldschmidt} V. M. Goldschmidt, Naturwissenschaften {\bf 14}, 477 (1926).

\bibitem{Shannon} R. D. Shannon, Acta Crystallogr., Sect. A {\bf 32}, 751 (19769).

\bibitem{howard98} C. J. Howard and H. T. Stokes, Acta Crystallogr., Sect. B: Struct. Sci. {\bf 54}, 782 (1998);
                   {\bf 58}, 565 (2002).

\bibitem{amisi12} S. Amisi, E. Bousquet, K. Katcho, and Ph. Ghosez, Phys. Rev. B {\bf 85}, 064112 (2012).

\bibitem{Ghosez99} Ph. Ghosez, E. Cockayne, U. V. Waghmare, and K. M. Rabe, Phys. Rev. B {\bf 60}, 836 (1999).

\bibitem{Machado11} R. Machado, M. Sepliarsky, and M. G. Stachiotti, Phys. Rev. B {\bf 84}, 134107 (2011).

\bibitem{bilbao}
M. I. Aroyo, A. Kirov, C. Capillas, J. M. Perez-Matoj, and H. Wondratschek, Acta Cryst. {\bf A62}, 115 (2006);
M. I. Aroyo, J. M. Perez-Mato, D. Orobengoa, E. Tasci, G. de la Flor, A. Kirov,
Bulg. Chem. Commun. 43(2) 183-197 (2011); 
M. I. Aroyo, J. M. Perez-Mato, C. Capillas, E. Kroumova, S. Ivantchev, G. Madariaga, A. Kirov \& H. Wondratschek,
Z. Krist. 221, 1, 15-27 (2006); 
D. Orobengoa, C. Capillas, M.I. Aroyo, and J. M. Perez-Mato, J. Appl. Cryst. {\bf A42}, 820 (2009).

\bibitem{isotropy} H. T. Stokes, D. M. Hatch, and B. J. Campbell, (2007).
                    ISOTROPY, stokes.byu.edu/isotropy.html.
\bibitem{woodward97}  P. M. Woodward, Acta Crystallographica {\bf B53}, 44 (1997).

\bibitem{Lufaso01} M. W. Lufaso and P. M. Woodward, Acta Crystallographica {\bf B57}, 725 (2001).         


\bibitem{rabe13} K. M. Rabe, in Functional Metal Oxides: New Science and Novel Applications,
                          ed. by Satish Ogale and V. Venkateshan, Wiley-VCH, Welnheim (2013).                                
    
\bibitem{darlington99}  C. N. W. Darlington and K. S. Knight, Acta Crystallographica {\bf B55}, 24 (1999).     

 \bibitem{bushmeleva} S. N. Bushmeleva, V. Yu, Pomjakushin, E. V. Pomjakushina, D. V. Sheptyakov,
                     A. M. Balagurov,
                     J. Magn. Magn. Mater. {\bf 305}, 491 (2006).
                      
\bibitem{jones} C. W. Jones, P. D. Battle, and P. Lightfoot,
                Acta Crystallogr., Sect. C: Cryst. Struct. Commun. {\bf 45}, 365 (1989).

\bibitem{yamaura} K. Yamaura and E. Takayama-Muromachi,
                  Phys. Rev. B  {\bf 64}, 224424 (2001).

\bibitem{nohara} Y. Nohara, S. Yamamoto, and T. Fujiwara,
                 Phys. Rev. B {\bf 79}, 195110 (2009).

\bibitem{knizek} K. Kn\'{\i}\v{z}ek, Z. Jir\'{a}k, J. Hejtm\'{a}nek, P. Nov\'{a}k, and W. Ku,
                 Phys. Rev. B {\bf 79}, 014430 {2009}.              


\bibitem{shannon} R. D. Shannon and C. T. Prewitt,
                  Acta Crystallogr. Sec. B {\bf 25}, 925 (1969).           

\bibitem{Alkauskas10}
A. Alkauskas, P. Broqvist, S. Pasquarello,
Phys. Status solidi (b) {\bf 248}, 775 (2010).
           
\bibitem{Rondinelli.AM.2012}   J. M. Rondinelli and C. J. Fennie,
             Adv. Mater. {\bf 24},  1961 (2012).   
   
\bibitem{Mulder.AFM.2013}   A. T. Mulder, N. A. Benedek, J. M. Rondinelli, and C. J. Fennie,
              Adv. Funct. Mater. {\bf 23}, 4810 (2013).                
                 
\end{thebibliography}
\end{document}